\documentclass[sigconf,9pt]{acmart}

\usepackage{color}
\usepackage{url}
\usepackage{booktabs}
\usepackage{multirow}
\usepackage{adjustbox}
\usepackage{listings}

\setcopyright{none}
\begin{document}

\title[Analyzing Vector Register Usage in Linux Packages to Understand Real-World Impact of Downfall Attack]{Analyzing Vector Register Usage in Linux Packages to\\Understand Real-World Impact of Downfall Attack}

\begin{abstract}
Downfall is a side-channel attack that leaks values in vector registers from a process to another on the same CPU core.
This attack enables an attacker to achieve serious outcomes (e.g., stealing AES keys), and there is no fundamental countermeasure besides applying microcode-based hardware patches.
Although the impact of this attack is discussed by the original paper and by Intel to some extent, it is still unclear whether programs used in daily computing activities of normal users are affected by Downfall. 
This paper thoroughly analyzes the usage of vector registers in widely used applications to assess the impact of Downfall on them.
In particular, we collect all packages (over 133~K) provided by the four latest long-term support versions of Ubuntu and measure various metrics on vector instructions.
Our findings include that over 60\% of all binary files contained in the packages use at least one vector register,
and that some highly popular packages such as apt might also be affected by Downfall.
\end{abstract}

\author{Yohei Harata}
\authornote{Currently in industry, work done while at Ritsumeikan University.}
%\email{s-akym@fc.ritsumei.ac.jp}
\affiliation{%
  \institution{Ritsumeikan University}
  \city{Ibaraki}
  \state{Osaka}
  \country{Japan}
}

\author{Soramichi Akiyama}
\affiliation{%
  \institution{Ritsumeikan University}
  \city{Ibaraki}
  \state{Osaka}
  \country{Japan}
}

\maketitle

\section{Introduction}
Side-channel attacks bypass normal communication channels to read or write data that the attacker is not supposed to do so.
They pose significant threats to security because (1) they can bypass conventional security measures such as password-based access control and
(2) they leave little to no trace of attack because a typical trigger of a side-channel attack is an execution of a program that looks normal.
For example, Spectre~\cite{Kocher2019} allows a process to steal data from other processes including the OS by abusing the speculative execution of the processor and a timing-based side-channel.
Another prominent example is RowHammer~\cite{Kim2014}, which allows a process to overwrite or even read data~\cite{Kwong2020} in a memory region of other processes including the OS by abusing an electromagnetic side-channel inside DRAM chips.

The Downfall attack~\cite{Moghimi2023} is a new side-channel attack found in 2023.
We refer to it simply as {\it Downfall} throughout this paper.
In some Intel CPUs, values in vector registers stored by a process can unintentionally leak to another process on the same core during speculative executions~\cite{Intel_GDS_technote}\footnote{Our descriptions of Downfall are based on the documentation from Intel~\cite{Intel_GDS_technote}, which does not necessarily coincide with the original paper of Downfall by Moghimi~\cite{Moghimi2023}.}.
This combined with other known attack primitives (e.g., cache-based timing side-channel) is a serious security threat.
An attacker is able to steal AES-256 keys from off-the-shelf OpenSSL, steal data processed by enclaves running on Intel SGX,
and even inject values into vector registers used by the victim process when conditions are met.

Although the applicability of Downfall is limited to vector registers, it is still a serious security threat.
First, vector registers are used not only for vector instructions (a.k.a. SIMD instructions) but also for scalar instructions.
This is because even a scalar floating point instruction (e.g., adding two single-precision floating point numbers) uses vector registers in x86\_64, where the usage of SSE is preferred over x87.
Second, vector instructions are generated by a compiler for performance optimization without the programmer's explicit notice.
Automatically generating vector instructions is of great interest in the compiler research area~\cite{Moghaddasi2025,Carpentieri2025,Adit2022}.

An unanswered question for Downfall is whether programs used in daily computing activities of normal users are affected or not.
The original paper~\cite{Moghimi2023} has revealed that a normal user-space process is capable of reading data from kernel memory or even from an Intel SGX enclave with Downfall.
However, it is unexplored whether applications used widely in daily computing activities are affected by Downfall.
To this end, we analyze the usage of vector registers in packages provided by Ubuntu, which is one of the most widely used Linux distributions.
We provide a wide range of analyses to understand the impact of Downfall,
such as the ratio of binary files that contain instructions vulnerable to Downfall and the lineage of these detected instructions.

Our analysis revealed important insights including:
\begin{enumerate}
%    \item Around 43\% of all packages in recent Ubuntu versions contain at least one binary file.
    \item Over 60\% of binary files in Ubuntu packages use vector registers, and the ratio monotonically increases as the Ubuntu version gets newer.
    \item Most of the vector registers are used by SSE instructions, but some binaries contain AVX instructions as well.
    \item Different versions of Ubuntu show different breakdowns of mnemonics for instructions using vector registers.
    \item Many usages of vector registers stem from shared libraries. Especially, the standard C and C++ libraries occupy a large portion of these libraries. 
    \item Some of the highly popular packages (e.g., \verb|apt|) also use vector registers even though they have no strong and obvious reason to do so.
\end{enumerate}

\section{Research Background}
\subsection{Vector Registers}
Vector registers in x86 and x86\_64 instruction set architectures (ISAs) are extensional registers that are either 128, 256, or 512-bit long.
More details of each type are given below:
\begin{enumerate}
    \item \verb|xmm| registers are 128-bit vector registers introduced in the Streaming SIMD Extensions (SSE) ISA.
    SSE defines eight registers (\verb|xmm0| to \verb|xmm7|), later extended to 16 (\verb|xmm8| to \verb|xmm15|).
    \item \verb|ymm| registers are 256-bit vector registers introduced in the Advanced Vector Extensions (AVX) ISA.
    AVX defines 16 registers (\verb|ymm0| to \verb|ymm15|), later extended to 32 (\verb|ymm16| to \verb|ymm31|).
    %The lower half of \verb|ymmN| is used as \verb|xmmN|, meaning that \verb|ymmN| and \verb|xmmN| cannot be used simultaneously for different purposes.
    \item \verb|zmm| registers are 512-bit vector registers introduced in AVX-512 as 32 new registers (\verb|zmm0| to \verb|zmm31|).
    %The lower half of \verb|zmmN| is used as \verb|ymmN|, meaning that \verb|zmmN| and \verb|ymmN| (and inductively \verb|xmmN|) cannot be used simultaneously for different purposes.
\end{enumerate}

Vector registers hold operands for two types of instructions, namely vector instructions and floating point instructions.
{\bf Vector instructions}, also known as SIMD (Single Instruction Multiple Data) instructions, apply the same operation to multiple data simultaneously
and are useful for applications with data-level parallelism, such as image processing and large-scale data analytics.
Vector registers store operands for vector instructions to hold multiple data simultaneously.
For example, \verb|zmm0| can hold up to 16 integers (32 bits each) to be used as the operand of vector addition.
{\bf Floating point instructions}, in contrast, can be either vector or scalar.
SSE {\it facilities for handling packed and {\bf scalar} single precision floating-point values contained in 128-bit registers} (Section 10.1 in ~\cite{Intel_arch_manual_vol1}, emphasized with {\bf bold fonts} by the authors of this paper).
This means even the operands of scalar floating point operations can be held by vector registers.
For example, gcc version 13.3.0 (Ubuntu 13.3.0-6ubuntu2-24.04) on Intel Core i5-11400 compiles the C code in Figure~\ref{figure:C_code} into the binary code in Figure~\ref{fig:binary_floating_point}.
We applied \verb|-O0| so that simple operations will not be optimized out.
We used \verb|objdump -D| to disassemble both the code and the data.
As in Figure~\ref{fig:binary_floating_point}, simple scalar operations on floating point numbers are served by \verb|xmm|.
We observed an equivalent result (only the addresses were different) with gcc version 12.2.0 (Debian 12.2.0-14+deb12u1) on Intel Xeon Platinum 8276.

\begin{figure}[t]
\begin{lstlisting}
float pi = 3.14;
float x = pi * 2.0;
\end{lstlisting}
\caption{\label{figure:C_code}Simple C Code of Floating Point Operations}
\end{figure}

\begin{figure}[t]
    \centering
    \includegraphics[width=\columnwidth]{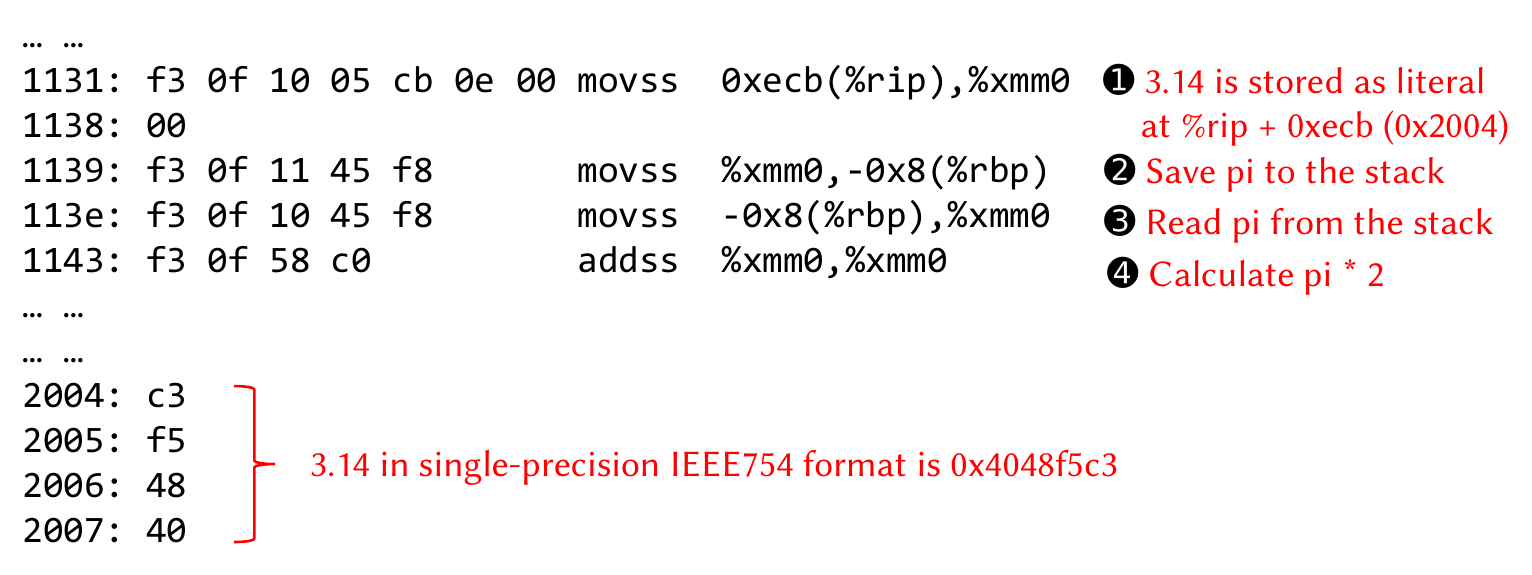}
    \caption{Binary Code Compiled from C Code in Figure~\ref{figure:C_code}}
    \label{fig:binary_floating_point}
\end{figure}

\subsection{The Position of This Paper on Downfall}
Downfall was first discovered by Moghimi~\cite{Moghimi2023}, and more detailed technical documentation was later provided by Intel~\cite{Intel_GDS_technote}.
Throughout this paper, we refer to the paper by Moghimi as the {\it original paper} and the documentation by Intel as the {\it Intel documentation}.
Based on these information sources, this paper takes the position that the Intel documentation is correct over the original paper whenever they disagree with each other.
This is because (1) the Intel documentation was published after the original paper
and (2) Intel has access to information not publicly available, such as the internal logic of their CPUs, to ensure the correctness of their documentation.

Specifically, there are two major differences between the Intel documentation and the original paper.
For the {\bf source of the data leakage}, the Intel documentation describes it as the vector registers themselves while it is suspected as a temporal buffer in the CPU in the original paper. 
The Intel documentation states that {\it stale data from previous usage of architectural or internal vector registers may get transiently forwarded to dependent instructions without being updated by the gather loads}~\cite{Intel_GDS_technote},
while the original paper claims that a temporal buffer transiently forwards data to later dependent instructions.
For the {\bf affected instructions}, the Intel documentation says that Downfall {\it expose data processed by instructions that use architectural vector registers explicitly ... as well as those that use internal vector registers implicitly}~\cite{Intel_affected_processors}.
On the other hand, the original paper lists specific instructions that are affected based on the experimental results (Table 2 in~\cite{Moghimi2023}).

\subsection{Details of Downfall}
{\bf Threat Model:}
We elaborate on the concrete threat model of Downfall.
It requires neither the root privilege nor the ability to execute suspicious instructions (e.g., \verb|clflush|).
\begin{enumerate}
    \item The attacker and the victim share the same physical machine. A typical case is that the attacker and the victim log in to a shared server via \verb|ssh|.
    \item The CPU of the shared physical machine is vulnerable to Downfall, and the microcode to fix it has not been applied. The list of CPUs vulnerable to Downfall can be found in~\cite{Intel_affected_processors}.
    \item The attacker can create a user-space program on the shared machine. For example, the attacker can type code on a text editor and compile it with \verb|gcc|, or they can transfer a pre-compiled binary via \verb|scp|.
    \item The attacker can execute the program created in (3) on the same physical core on which the victim's process is running with a user-level privilege.
    Even when the attacker has no explicit control on the core to execute their program,
    the attacker can invoke the same program many times in the hope that it will eventually be executed on the same core.
\end{enumerate}

\begin{figure}[t]
    \centering
    \includegraphics[width=0.95\columnwidth]{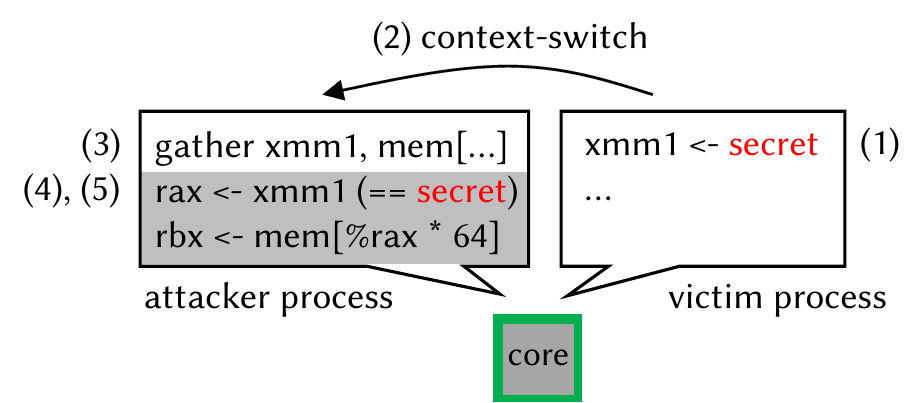}
    \caption{Procedures of Downfall Attack}
    \label{fig:procedure_downfall}
\end{figure}

{\bf Procedures of Data Leakage:}
Under the above threat model, the attacker process can leak data from the victim process as follows.
The procedures are also illustrated in Figure~\ref{fig:procedure_downfall}.

\begin{enumerate}
\item The victim process $P_V$ stores a value to a vector register $R$ (e.g., \verb|xmm1|) as a result or as an operand of a vector instruction or a floating point instruction.
\item The OS decides to preempt $P_V$, and the core on which $P_V$ has been executed is given to the attacker process $P_A$.
\item $P_A$ executes a \verb|gather| instruction with $R$ specified as the destination register.
This instruction tries to read values from multiple non-consecutive addresses in memory and stores the merged results into the vector register $R$.
\item The CPU enters the speculative execution mode because \verb|gather| incurs a large delay due to random memory accesses.
Instructions with gray background in Figure~\ref{fig:procedure_downfall} are executed in the speculative execution mode.
The attacker increases the chance and the duration of the speculative execution by (i) specifying uncacheable memory locations to \verb|gather| and (ii) loading another uncacheable address to a normal register by a \verb|mov| instruction right before \verb|gather|.
\item During the speculative execution, {\it stale data from previous usage of architectural or internal vector registers may get transiently forwarded to dependent instructions without being updated by the gather loads}~\cite{Intel_GDS_technote}.
This means that instructions issued by $P_A$ subsequent to the \verb|gather| instruction can observe the value stored to $R$ by $P_V$.
\end{enumerate}

{\bf Materialization of Transient Value:}
Since the value leaked to the attacker process is only visible during speculative execution\footnote{If it is still visible after the speculative execution, that is a violation of the ISA spec because the result of gather observed by the attacker process is simply wrong.}, the attacker needs {\it materialization} of the value.
It means to (i) convert the transient value to an architectural state that is not squashed even after the end of the transient execution,
and (ii) observe the architectural state after the speculative execution to recover the transient value.
A typical method to achieve this is to conduct a Prime+Probe~\cite{Liu2015} attack on itself as follows.

\begin{enumerate}
    \item Before conducting Downfall, the attacker process occupies all the last-level cache lines with its data by accessing a large array.
    This corresponds to the prime phase of Prime+Probe.
    \item The attacker process conducts Downfall and copies the leaked transient value to a normal register such as \verb|rax|.
    This copy is needed to use the leaked transient value as an address later because xmm registers {\it cannot be used to address memory} (Section 10.2.2 in~\cite{Intel_arch_manual_vol1}).
    Note that the value in \verb|rax| is still transient and going to be squashed later.
    \item The attacker process accesses the address 64 $\times$ \verb|%rax| in its own memory address space, where \verb|%rax| indicates the value stored in the \verb|rax| register and 64 is the size of a cache line.
    This evicts a cache line at the \verb|%rax|$^{\rm th}$ set\footnote{We assume that the number of sets is larger than \%rax for simplicity.}.
    Now that the transient value has been converted to an architectural state because the contents of the cache are not squashed even after a mispredicted speculative execution.
    \item The attacker process accesses the large array it has accessed in (1) again.
    By measuring the access latencies, the attacker process can detect which part of the array is evicted (which cache set is used) and thus infer the leaked value.
    This corresponds to the probe phase of Prime+Probe.
\end{enumerate}
Note that any variant of cache-based timing side-channel besides Prime+Probe can be applied here.
For example, the original paper suggests using Flush+Reload~\cite{Yarom2014}.

\section{Impact of Downfall}
\subsection{\label{section:existing_discussions}Existing Discussions}
The impact of Downfall is already discussed both by the original paper and by Intel to some extent.
The original paper revealed many concerning attack scenarios.
It shows that an attacker can (i) steal AES round keys from OpenSSL that uses the AES-NI instructions,
(ii) steal data from a victim process that has not yet even accessed the stolen data (which is made possible due to the CPU prefetching),
(iii) steal data from the Linux kernel when kernel code (e.g., a kernel module) uses \verb|memcpy| that is compiled into binary with internal uses of vector registers, and
(iv) even inject data to vector registers used by a victim process if the victim code has a \verb|gather| instruction (i.e., exploiting Downfall in the other way around).

%\paragraph{Discussion by Intel}
The threat analysis guidance~\cite{Intel_threat_anaylsis} by Intel discusses the potential exposure of different systems to Downfall.
It categorizes systems by whether they are single-user or multi-tenant, and whether they run trusted code only or untrusted code as well.
Their takeaways include that (i) a system running untrusted code has higher exposure to Downfall and
(ii) even a single-user system such as a high-performance computing system with strong user isolation could be at risk when the login or interactive nodes are shared across multiple users, which is often the case.

\subsection{Impact on Daily Computing Activities}
The existing discussions on the impact of Downfall lack consideration for daily computing activities.
Here, a {\it daily computing activities} refer to tasks that a normal user executes on a computer at work or home,
such as writing documents on a text editor and updating the installed software on a package manager.
The lack of consideration stems from the fact that the original paper is too specific and the Intel documentation is too generic.
The original paper shows particular attack scenarios such as stealing AES keys from OpenSSH, but it does not investigate broader applications.
On the other hand, the Intel documentation discusses in a top-down manner by categorizing the entire systems as explained previously, but does not mention the exposure of specific applications to Downfall.

Given the lack of consideration of the impact of Downfall on daily computing activities,
this study investigates the usage of {\it target instructions} that are affected by Downfall in a wide range of applications.
The definition of target instructions is given in Section~\ref{section:target_instructions}.
Specifically, we investigate applications provided as packages by the official software repository of Ubuntu.
Ubuntu is a widely used Linux distribution and is ranked 6$^{\rm th}$ in the 2024 page hit rankings of DistroWatch~\cite{distrowatch}, which compiles information on various Linux distributions.
Linux Mint and Debian GNU/Linux (ranked 2$^{\rm nd}$ and 4$^{\rm th}$ in the same ranking, respectively) are also in the same family of distributions\footnote{Linux Mint is based on Ubuntu and Ubuntu is based on Debian GNU/Linux.}.
These facts lead us to define Ubuntu packages as representative of applications used in daily computing activities and thus the subject of our study.

\section{Quantitative Analysis}
\subsection{\label{section:target_instructions}Target Instructions}
We define the {\it target instructions} of our analysis as machine instructions that meet either of the following conditions:

\begin{enumerate}
    \item Instructions that explicitly use vector registers.
    These include vector instructions such as \verb|movdqa| that copies multiple integers at once,
    and floating point instructions with either vector or scalar operands such as \verb|addss| that adds two single-precision floating point numbers. 
    \item Instructions that implicitly use vector registers.
    They internally use vector registers for performance optimization when combined with the \verb|rep| prefix.
    Although the \verb|rep| prefix can be used with many instructions, we specifically target \verb|movs|, \verb|movsb|, \verb|movsw|, \verb|movsd|, and \verb|movsq|.
    This is because the Intel documentation only mentions the \verb|movs| family as the example of this category,
    even though other instructions with the prefix might use vector registers internally.
\end{enumerate}

Our definition of the target instruction aligns with the Intel documentation.
Specifically, our target instructions include ones that only take registers as their operands (e.g., \verb|vmovdqa xmm1 xmm2|).
These instructions are not included in the analysis of the original paper as the original paper suspects that Downfall is applicable only when one of the operands is a memory location.

\subsection{Analysis Metrics and Targets}
The specific metrics we measure to assess the impact of the Downfall are as follows:
\begin{enumerate}
    \item Ratio of binary files containing target instructions among all binary files in all available packages.
    The definition of a {\it binary file} is given later.
    \item Ratio of target instructions from each ISA (e.g., SSE) among all the detected target instructions.
    \item Ratio of each mnemonic (e.g., \verb|movss|) among all the detected target instructions.
    \item Ratio of target instructions originating from shared libraries among all target instructions whose origins are known.
    \item Ratio of target instructions from each shared library.
\end{enumerate}
We also conduct individual analyses of popular packages to draw more concrete findings.

We measure these metrics on the 64-bit flavors of four LTS (Long-Term Support) Ubuntu versions, namely 24.04 LTS, 22.04 LTS, 20.04 LTS, and 18.04 LTS.
The LTS versions are supported for 10 years, which is much longer than other versions.
An LTS version is released every other year.

\subsection{Analysis Methodology}
\begin{figure}[t]
    \centering
    \includegraphics[width=0.95\columnwidth]{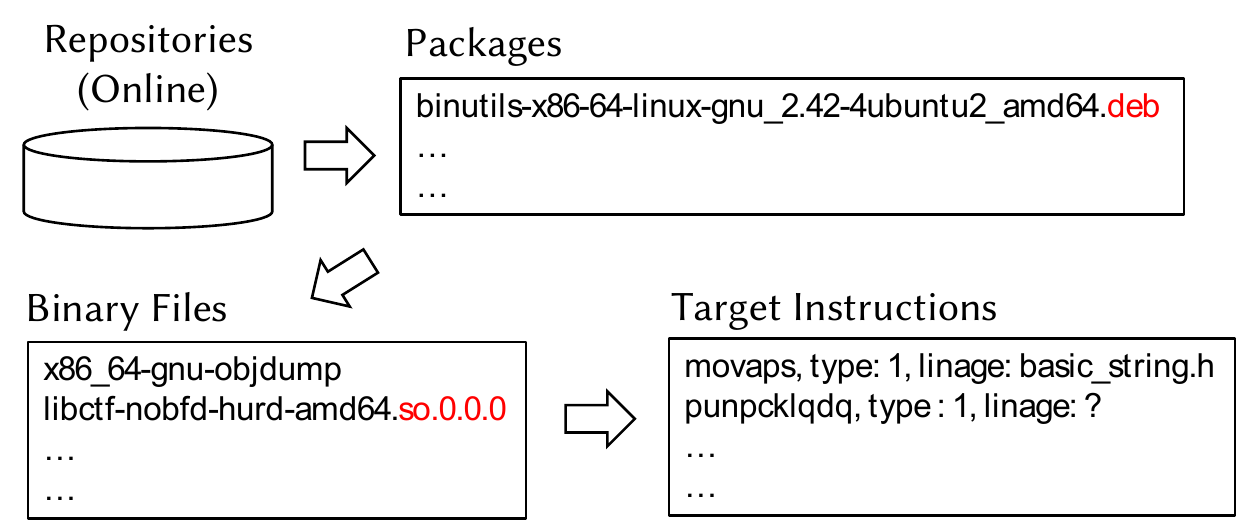}
    \caption{Analysis Methodology}
    \label{fig:methodology}
\end{figure}

Ubuntu provides applications to users as packages.
A {\it package} contains the application itself, the manual, and additional metadata that describes how the it should be installed and the dependencies on other packages.
The application itself can be a set of ELF files when it is developed with a compiler-based language (e.g., C, C++),
or can be provided as a text-based programs when it is written in scripting languages (e.g., Python, Ruby).

The methodology of our analysis is as follows and illustrated in Figure~\ref{fig:methodology}.
Each procedure is explained later in detail.

\begin{enumerate}
    \item Collect all packages for each Ubuntu version.
    \item Extract binary files from the collected packages.
    \item Identify target instructions from each binary file.
    \item Determine the lineage of each target instruction.
\end{enumerate}

{\bf Collect all packages:} We collect all packages provided in each Ubuntu version using the \verb|apt-mirror| command.
This command downloads all packages for a specified repository and Ubuntu version.
We download packages from all four repositories (\verb|main|, \verb|universe|, \verb|multiverse|, and \verb|restricted|) and save them in the archive format without installation.
The differences between these repositories are described in the official wiki~\cite{ubuntu_help_repositories}.
Table~\ref{table:num_of_packages_and_binaries} shows the number of collected packages for each Ubuntu version.
The total size of the collected packages is around 410~GB.

\begin{table}[t]
\centering
\caption{Number of Collected Packages and Binary Files}
\label{table:num_of_packages_and_binaries}
 \begin{tabular}{rrrrr}
 \hline
  & 24.04 LTS   & 22.04 LTS     & 20.04 LTS    & 18.04 LTS \\
  \hline
  Packages & 37,827  & 34,745  & 31,253  & 29,611 \\
  Binaries  & 117,122 & 104,380 & 100,784 & 106,986 \\
 \hline
 \end{tabular}
%\end{center}
\end{table}

{\bf Extract binary files:} After all the packages are downloaded, we extract binary files from them.
Table~\ref{table:num_of_packages_and_binaries} shows the number of binary files extracted for each Ubuntu version.
A binary file is a file that meets one of the following conditions.
\begin{enumerate}
    \item The execute permission is enabled and the file type given by \verb|file -i| command is either \verb|application/x-executable| or \verb|application/x-pie-executable|.
    \item The filename ends either with \verb|.so(.[digits])*| or \verb|.a|.
    The notation \verb|(x)*| indicates zero or more repetitions of \verb|x| and \verb|[digits]| indicates one or more repetitions of a number.
    For example, \verb|liba.so| and \verb|liba.so.1.2| meet this condition.
\end{enumerate}

The intention of each condition is as follows.
Condition (1) finds executable files but excludes text-based scripts (e.g., \verb|.py| files) by checking the file type.
Checking the execution permission speeds up the process; running \verb|file -i| on all files in all packages is too time-consuming.
Condition (2) finds dynamically linked libraries.
They contain binary instructions but lack the execution permissions, thus need to be identified by their filenames.

{\bf Identify target instructions:}
Each instruction in the extracted binary files is investigated to determine if it is a target instruction.
We first use the \verb|objdump| command with the \verb|-d| option to extract instructions but not anything else (e.g., ELF header, data section).
We then input each line of the output from \verb|objdump| to a disassembler named Zydis~\cite{zydis},
which can identify the mnemonic and the operands (e.g., \verb|xmm| register) of a piece of binary code.
We distinguish the two types of target instructions (either use vector registers explicitly or implicitly) from the disassembling results.

{\bf Determine the lineage:}
The lineage of each instruction is investigated to determine if it comes from a shared library.
We use the debug information of each package provided by a dedicated repository~\cite{ubuntu_ddeb}.
The debug information may (but not necessarily) contain the file path and the line number of the source code that an instruction originates from.
An instruction is determined to be from a shared library if the corresponding file path starts either with \verb|/usr/include| or \verb|/usr/lib|.
The limitation of this method is that it could lead to false negatives when the developer of a package has copied a shared library in the source tree of the package itself.

We determine the lineage of instructions only for Ubuntu 24.04 LTS (the latest version we use in this study).
This is because the repositories providing the symbol information are not mirrored and quite unreliable, unlike the repositories for the application packages.
Due to this reason, we inserted a wait-time of 1 second after downloading the symbol information for each package, resulting in several hours to download all the provided symbol information.
We refer to this limitation as {\it the data availability reason} hereafter.

\section{Results of Anaylsis}
\subsection{Ratio of Packages with Binary Files}
Table~\ref{table:ratio_of_package_with_binary} shows the ratio of packages that contain at least one binary file among all the packages.
Examples of the packages with no binary files include ones that consist of programs written in scripting languages (e.g., \verb|a2d| consists of Python and Shell scripts), and ones that contain documentation of another package (e.g., \verb|alpine-doc|).
We count packages \verb|xxx| and \verb|xxx-doc| separately when they are provided as independent \verb|.deb| files.
The result shows that the ratio is fairly consistent among all the Ubuntu versions we investigated.

Packages that do not contain any binary files (i.e., the other 57~\%) are not affected by Downfall {\it by themselves}.
However, they could also be affected through other packages that they depend on.
For example, a program written in a scripting language is merely a text file, but it can be affected by Downfall if the runtime of the language (e.g., an interpreter) contains target instructions.

\begin{table}[h]
%\begin{center}
\centering
\caption{Ratio of Packages Containing Binary Files}
\label{table:ratio_of_package_with_binary}
 \begin{tabular}{rrrr}
 \hline
  24.04 LTS   & 22.04 LTS     & 20.04 LTS    & 18.04 LTS \\
  \hline
  43.2 \% & 42.7 \%   & 43.2 \%  & 43.1 \% \\
 \hline
 \end{tabular}
%\end{center}
\end{table}

\subsection{Ratio of Binary Files Containing Target Instructions}
\label{section:ration_of_binary_with_target_insts}

\begin{figure}[h]
    \centering
    \includegraphics[width=0.95\columnwidth]{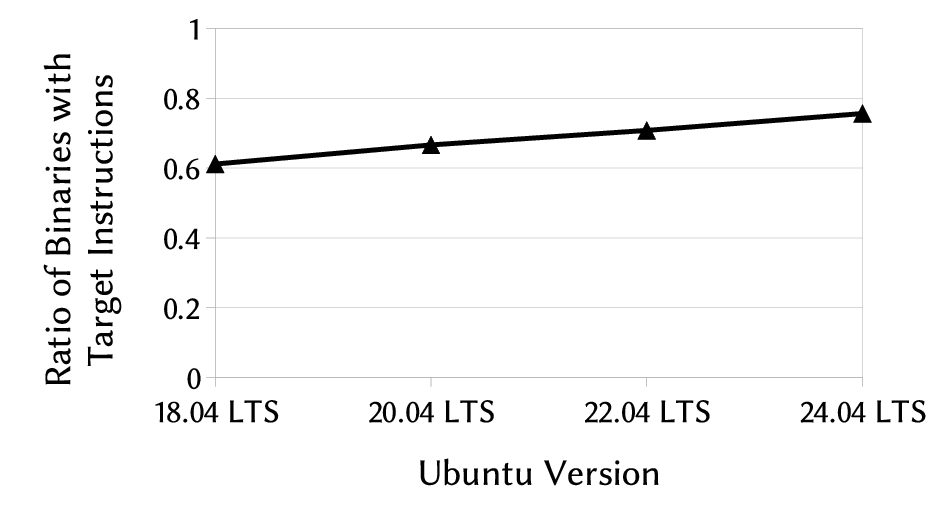}
    \caption{Ratio of Binary Files Containing Target Instructions}
    \label{fig:ratio_binaries}
\end{figure}

Figure~\ref{fig:ratio_binaries} shows the ratio of binary files containing target instructions relative to all binary files for each Ubuntu version.
We make the following observations from the result:
\begin{enumerate}
    \item Over 60\% of all the binary files in all investigated Ubuntu versions contain target instructions.
    \item The ratio of binary files containing target instructions increases monotonically as the Ubuntu version gets newer.
\end{enumerate}

We hypothesize that the monotonic increase we observe in the ratio stems from the increasing ability of the compiler to generate highly optimized code with vector instructions.
For example, the default C/C++ compiler of Ubuntu 24.04 LTS is gcc 14~\cite{ubuntu_24_release_notes},
which newly {\it supports vectorizing loops which contain any number of early breaks}~\cite{gcc14_release_notes}.

\subsection{Ratio of Target Instructions from Each ISA}
\begin{figure}[t]
    \centering
    \includegraphics[width=0.9\columnwidth]{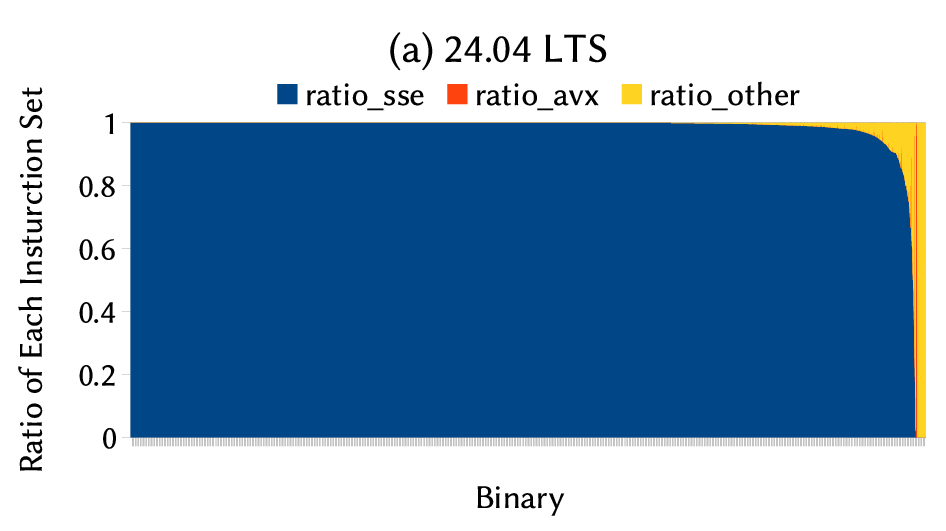}
    \includegraphics[width=0.9\columnwidth]{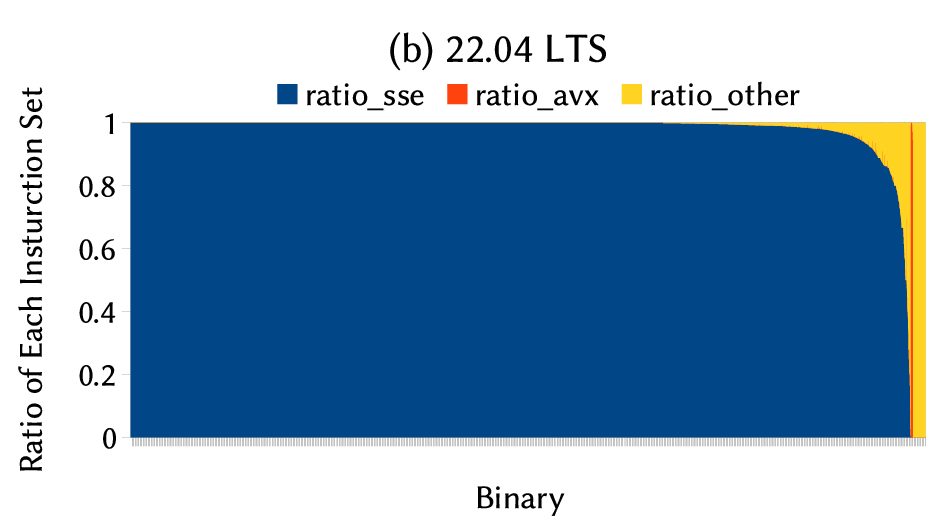}
    \includegraphics[width=0.9\columnwidth]{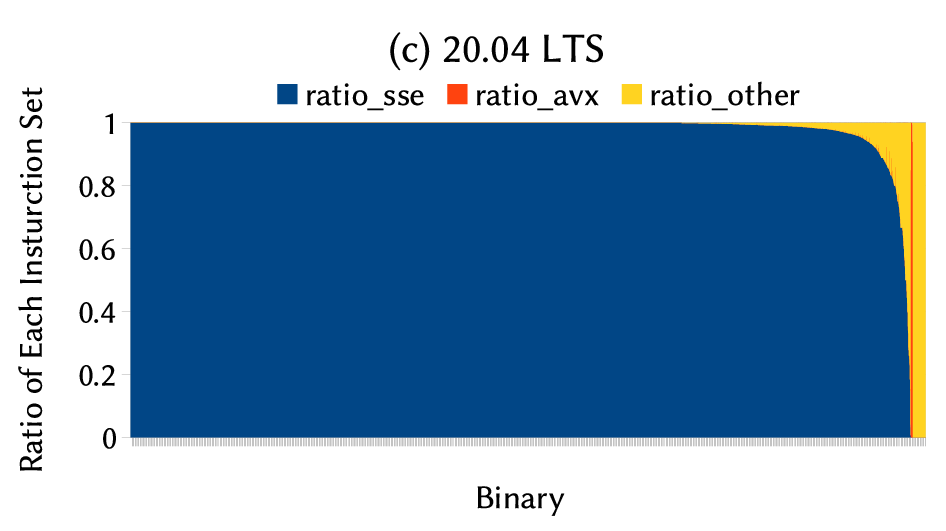}
    \includegraphics[width=0.9\columnwidth]{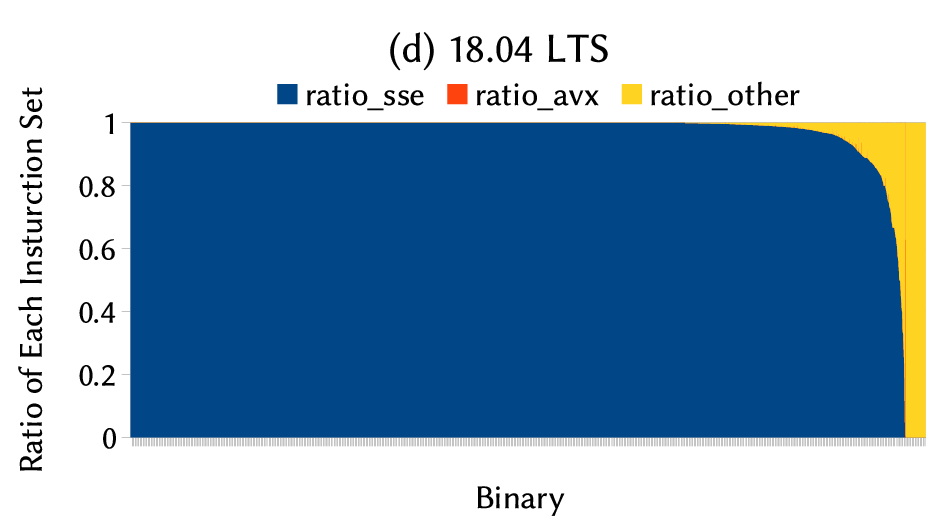}
    \caption{Ratio of Target Instructions by ISAs}
    \label{fig:instructions_per_isa}
\end{figure}

Figure~\ref{fig:instructions_per_isa} shows the ratio of target instructions from each ISA (e.g., SSE) among all the detected target instructions in each binary file.
The x-axis indicates binary files and the y-axis shows the ratio of target instructions from each ISA with different colors.
The binary files are sorted in descending order of \verb|ratio_sse|.
The details of each category are as follows:
\begin{enumerate}
    \item \verb|ratio_sse|: Instructions included SSE, SSE2, SSE3, and SSE4 are counted as this category.
    \item \verb|ratio_avx|: Instructions included in AVX, AVX2, and AVX-512 are counted as this category.
    \item \verb|ratio_other|: Instructions included in neither SSE nor AVX but use vector registers internally when combined with the \verb|rep| prefix are counted as this category.
    The detailed definition of these instructions is given in Section~\ref{section:target_instructions}.
\end{enumerate}

We make the following observations from the result:
\begin{enumerate}
    \item Most of the target instructions are from the SSE ISA.
    This is because floating point operations are normally served by SSE rather than by legacy x87 in 64 bit x86 CPUs.
    \item Some binary files show very large \verb|ratio_avx|, prominent as spikes in the graph.
    \item The occupancy of \verb|ratio_sse| in the graphs monotonically increases as the Ubuntu version gets newer.
    This means that the use of \verb|rep movs*| instructions is less and less common in newer Ubuntu versions.
\end{enumerate}

\subsection{Ratio of Each Mnemonic}
\begin{table}[t]
\centering
\caption{Ratio of Target Instructions by Mnemonic (\%)}
\label{table:ratio_of_each_mnemonic}
 \begin{tabular}{rrrr}
 \hline
  24.04 LTS   & 22.04 LTS     & 20.04 LTS    & 18.04 LTS \\
  \hline
  movups (12)   & movsd (14)    & movsd (15)   & movsd (25) \\
  movaps (11)   & movaps (8.3)  & movaps (8.2) & movss (8.5) \\
  movsd (10)    & movups (7.9)  & movss (7.2)  & movaps (7.5) \\
  movq (5.2)    & movss (5.7)   & movups (6.6) & movups (5.7) \\
  movdqa (5.1)  & {\bf mulsd} (3.5)   & {\bf mulsd} (3.4)  & {\bf mulsd} (5.5) \\
  movdqu (3.8)  & vmovaps (3.2) & movapd (2.7) & {\bf pxor} (4.6) \\
  {\bf pxor} (3.7)    & movdqa (3.0)  & {\bf pxor} (2.6)   & movapd (4.1) \\
  movss (3.7)   & {\bf pxor} (2.9)    & movdqa (2.5) & {\bf addsd} (3.9) \\
  {\bf mulsd} (2.6)   & movdqu (2.9)  & movdqu (2.4) & movdqa (3.5) \\
  movapd (2.2) & movapd (2.5)  & {\bf addsd} (2.3)  & movdqu (2.7) \\
 \hline
 \end{tabular}
\end{table}

Table~\ref{table:ratio_of_each_mnemonic} shows the ratio of each mnemonic among all the detected target instructions for each Ubuntu version.
The numbers in parentheses represent the percentage relative to all target instructions in the respective version.
We only show the top 10 instructions for brevity.
Instructions in {\bf bold} are those not in the \verb|mov| family.
The mnemonic \verb|movsd| is both in SSE and vanilla x86 (the latter uses vector registers internally when combined with the \verb|rep| prefix),
but all \verb|movsd| instances in Table~\ref{table:ratio_of_each_mnemonic} are from SSE.

We make the following observations from the result:
\begin{enumerate}
    \item Most of the target instructions are from the \verb|mov| family. These instructions are involved in data movement either between two vector registers or a vector register and memory.
    \item Some instructions such as \verb|pxor| and \verb|mulsd| perform computations on values on vector registers.
    \item In versions up to 22.04 LTS, \verb|movsd| has a significantly higher ratio than other mnemonics.
    However, this trend changes in 24.04 LTS, where \verb|movups| is the most prevalent.
    It is an important insight in the context of Downfall because \verb|movups| copies two double precision numbers instead of one copied by \verb|movsd|.
    This suggests that Downfall may be able to leak more data in one attempt in newer Ubuntu versions.
\end{enumerate}

\subsection{Ratio of Target Instructions from Shared Libraries}
\begin{figure}[t]
    \centering
    \includegraphics[width=0.95\columnwidth]{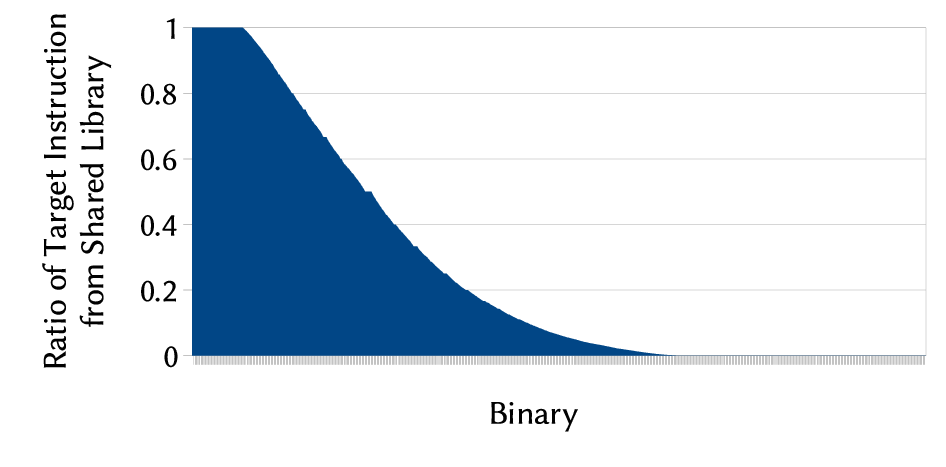}
    \caption{Ratio of Target Instructions from Shared Libraries among Those with Known Lineage}
    \label{fig:ratio_of_insts_from_library}
\end{figure}

Figure~\ref{fig:ratio_of_insts_from_library} shows the ratio of target instructions that originate from shared libraries among the ones with known lineage.
Suppose we have a binary file with three target instructions and the lineage of each instruction is (i) a share library, (ii) some other binary (e.g., \verb|mycode.o|), and (iii) unknown, respectively.
The ratio for this binary is calculated as $\frac{1}{2}$, whose denominator excludes the target instruction (iii).
The x-axis indicates binary files sorted in descending order of the y values.
This experiment is conducted only for Ubuntu 24.04 LTS due to the data availability reason.

We observe that target instructions in some binaries are all from shared libraries, while in other binaries they are all from non-library code.
This suggests that eliminating target instructions to avoid Downfall requires modifying both shared libraries and non-library code, depending on the locations of the target instructions.
%We investigate which libraries contain many target instructions in Section~\ref{section:ratio_of_each_library}.

\subsection{\label{section:ratio_of_each_library}Ratio of Target Instructions by Linage}
\begin{figure}[t]
    \centering
    \includegraphics[width=\columnwidth]{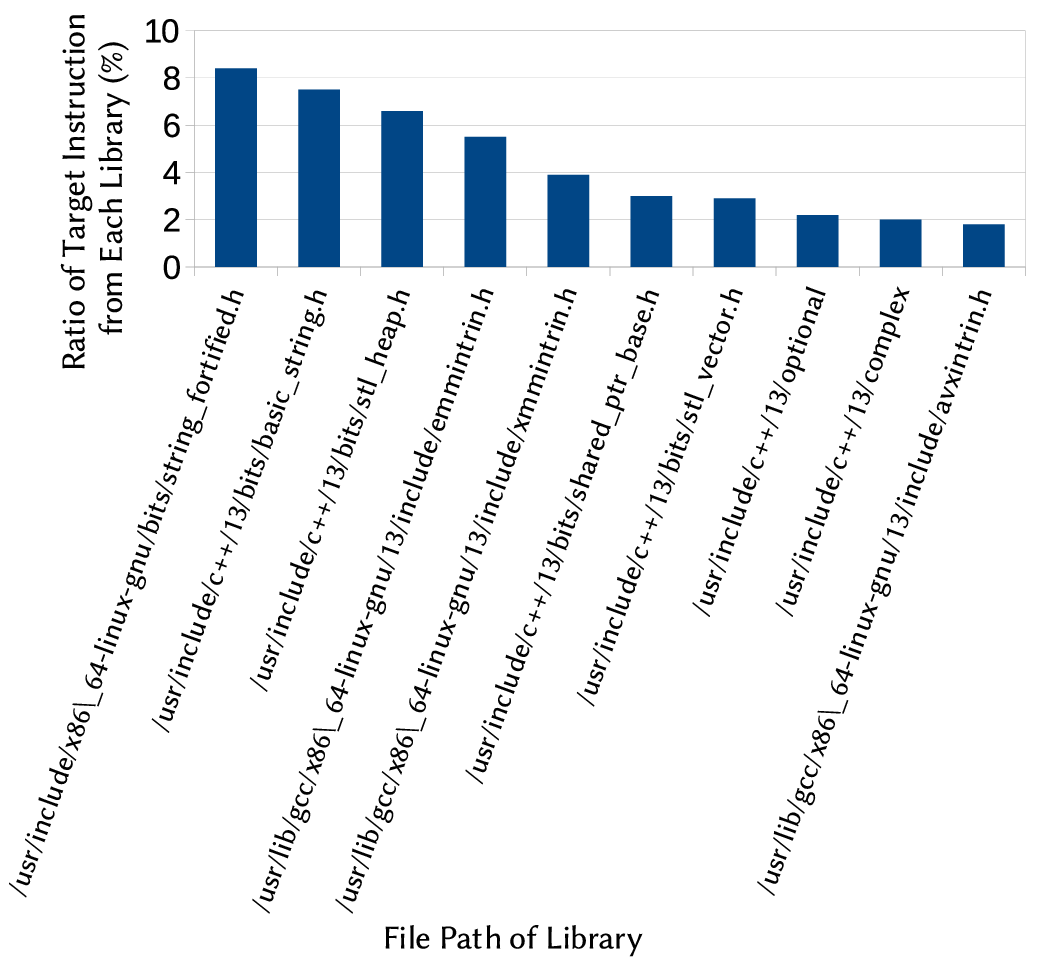}
    \caption{Ratio of Target Instructions from Each Library}
    \label{fig:ratio_of_each_library}
\end{figure}

Figure~\ref{fig:ratio_of_each_library} shows the ratio of target instructions from each shared library among those originating from shared libraries.
For example, 8.4~\% of the target instructions originating from shared libraries are from a single library named \verb|/usr/.../_fortified.h| (the intermediate paths are omitted for brevity).
This experiment is conducted only for Ubuntu 24.04 LTS due to the data availability reason.
%A library \( L \) is considered the source of an instruction \( i \) if the debug information indicates that \( i \)’s source code is included in \( L \). Note that Ubuntu’s debug information may only include symbol information, so not all vulnerable instructions have identifiable source libraries. 

We make the following observations from the result:
\begin{enumerate}
    \item The top 10 libraries shown in the figure occupy around 44~\% of all the known origins of the target instructions.
    \item The top 1 library (i.e., \verb|string_fortified|) is a secure version of conventional \verb|string.h|.
    This library utilizes a buffer overflow protection mechanism provided by the compiler as a form of built-in functions such as \verb|__builtin___memcpy_chk|.
    \item 6 out of the top 10 libraries (i.e., \verb|basic_string|, \verb|stl_heap|, \verb|shared_ptr_base|, \verb|stl_vector|, \verb|optional|, \verb|complex|) are implementations of standard libraries of either C or C++.
    \item 3 out of the top 10 libraries (i.e., \verb|emmintrin|, \verb|xmmintrin|, and \verb|avxintrin|) enable a programmer to explicitly tell the compiler to generate vector instructions.
    Their only use case is performance optimization as far as we know.
\end{enumerate}

The observations (2) and (3) imply that many target instructions are generated by the compiler without the programmer's notice.
This is beneficial for compute-bound programs such as data compression and video editing tools (e.g., \verb|gzip|, \verb|ffmpeg|, etc).
However, many daily-use programs are (1) either IO- or human-bound (e.g., \verb|dpkg|) or (2) of little interest in terms of their performance (e.g., \verb|passwd|) as we analyze in Section~\ref{section:per_package_analysis}.
In these cases, generating vector instructions without the notice of the programmer opens up an attack surface to Downfall while providing no benefit.

\if 0
\begin{table}[t]
\caption{Proportion of vulnerable instructions by source library (PROBABLY MAKE IT A GRAPH, RATHER THAN A TABLE)}
\label{table:ratio_of_each_library}
\begin{center}
 \begin{tabular}{|c|l|c|}
 \Hline %% ←
  Rank & File Path & Proportion \\ \hline
  1 & /usr/include/x86\_64-linux-gnu/bits/string\_fortified.h & 8.4 \% \\
  2 & /usr/include/c++/13/bits/basic\_string.h & 7.5 \% \\
  3 & /usr/include/c++/13/bits/stl\_heap.h & 6.6 \% \\
  4 & /usr/lib/gcc/x86\_64-linux-gnu/13/include/emmintrin.h & 5.5 \% \\
  5 & /usr/lib/gcc/x86\_64-linux-gnu/13/include/xmmintrin.h & 3.9 \% \\
  6 & /usr/include/c++/13/bits/shared\_ptr\_base.h & 3.0 \% \\
  7 & /usr/include/c++/13/bits/stl\_vector.h & 2.9 \% \\
  8 & /usr/include/c++/13/optional & 2.2 \% \\
  9 & /usr/include/c++/13/complex & 2.0 \% \\
  10 & /usr/lib/gcc/x86\_64-linux-gnu/13/include/avxintrin.h & 1.8 \% \\
 \Hline %% ←
 \end{tabular}
\end{center}
\end{table}
\fi

\subsection{Detailed Analysis of Popular Packages}
\label{section:per_package_analysis}
\begin{table}[t]
\centering
\caption{Detailed Analysis Results of Popular Packages}
\label{table:analysis_popular_packages}
 \begin{tabular}{clrrl}
 \hline
  {\it R} & Package Name & $N_T$ &$N_F$ & $F_{\rm MAX}$ \\
  \hline
  1 & adduser & 0 & 0 & {\it N/A} \\
  2 & dpkg    & 2475 & 42 & {\bf ensure\_diversions} \\
  3 & debconf & 0 & 0 & {\it N/A} \\
  4 & apt     & 2938 & 150 & {\bf FTPConn::Open} \\
  5 & libacl1 & 35 & 4 & {\it M/F} \\
  6 & libbz2-1.0 & 377 & 3 & {\bf BZ2\_decompress} \\
  7 & libblkid1 & 695 & 29 & {\bf memcpy} \\
  8 & libpam-modules & 568 & 22 & {\bf pam\_sm\_authenticate} \\
  9 & passwd & 1040 & 23 & {\bf main} \\
  10 & tar & 1193 & 70 & {\bf get\_date\_or\_file} \\
 \hline
 \end{tabular}
\end{table}

Table~\ref{table:analysis_popular_packages} shows the detailed analysis results of popular packages.
The popularity is based on the number of installations according to the package popularity ranking~\cite{debian_package_ranking} of Debian GNU/Linux.
We use Debian's ranking because Debian and Ubuntu are closely related distributions, and no equivalent ranking exists for Ubuntu as far as we know.
This experiment is conducted only for Ubuntu 24.04 LTS due to the data availability reason.

{\it R} in the table shows the position in the package popularity ranking as of July 2025.
We use the {\it inst} ranking (based on the number of installations) from the {\it all reports} (counted for all Debian versions).
Note that the ranking may be slightly different from time to time as it is dynamically generated and no static ranking exists as far as we know.
$N_T$ and $N_F$ indicate the total number of target instructions detected in each package and the number of binary files in the package containing at least one target instruction, respectively.
$F_{\rm MAX}$ denotes the function name with the largest number of target instructions within each package,
with $N/A$ indicating there is no such function (as $N_T$ is 0) and ${\it M/F}$ indicating the target instructions are equally distributed to multiple functions.

We discuss the justification of using vector registers for each package, except for \verb|adduser| and \verb|debconf| (whose $N_T$ are 0).
\begin{description}
    \item [dpkg] mainly does two tasks on software packages compressed as a \verb|.deb| file: uncompressing it and copying extracted files to particular directories specified by config files.
    The copying task is IO-bound and there is no obvious and strong need to use vector registers for this task.
    The uncompressing task, however, is computational-bound and benefits from vector instructions using vector registers.
    \item [apt] is a front-end of \verb|dpkg| that automatically solves the dependencies and fetches required packages from a repository (mostly through a network).
    These additional features have no obvious and strong need to use vector registers.
    \item [libacl1] implements the access control list (ACL) feature defined in the POSIX standard. The implemented functions include \verb|acl_add_perm| that adds a new permission (either read, write, or execute) to an object, and \verb|acl_valid| that validates a given access control list.
    The list of all implemented functions is in~\cite{acl_manual}.
    There is no obvious and strong need to use vector registers because these functions do not require floating point instructions nor high performance. 
    \item [libbz2-1.0] implements a compression algorithm based on the Burrows-Wheeler transform and Huffman coding. 
    This algorithm benefits from parallelism achieved by vector instructions, thus there is a need to use vector registers.
    \item [libblkid1] helps system utilities such as \verb|mount| find a block device from its UUID or label.
    There is no obvious and strong need to use vector registers because this feature does not require floating point instructions nor high performance. 
    \item [libpam-modules] contains kernel modules to implement the Pluggable Authentication Modules (PAM) feature such as \verb|pam_unix.so|.
    There is no obvious and strong need to use vector registers because it does not require high performance, and the use of floating point numbers is prohibited inside the Linux kernel~\cite{Linux_FP}.
    \item [passwd] allows a user to change their login password. 
    There is no obvious and strong need to use vector registers because it does not require floating point instructions nor high performance.
    Note that hashing a given password can benefit from vector instructions, but it is done outside of this package by a library (e.g., \verb|libcrypt1|).
    \item [tar] archives multiple files into one and then compresses the archived file (also the other way around: uncompressing and then unarchiving).
    The compression and uncompression benefit from vector instructions using vector registers.
\end{description}

\section{Related Work}
\paragraph{Statistical Analysis of Binary Code}
Akshintala~{\it et al.}~\cite{Akshintala2019} analyzed the packages in Ubuntu 16.04 to understand the importance of different x86\_64 instructions.
Their findings include that \verb|mov| is the most popular instruction family and that around 98~\% of packages use \verb|movaps|.
Compared to~\cite{Akshintala2019},
(1) our analysis is more fine-grained because we analyze each binary file independently while their analysis is per-package, and
(2) we specifically focus on Downfall and count scalar instructions that internally use vector instructions differently from other scalar ones.
Priyadarshan~{\it et al.}~\cite{Priyadarshan2024} proposed a method to accurately disassemble binary files of complex ISAs such as x86 to help conduct binary analysis.
Their main idea is to distinguish code and data in the binary by leveraging statistical characteristics of them such as byte distribution.
Although our analysis can be more accurate by using their method (we use vanilla \verb|objdump|),
most of our results could be consistent because they are statistical except for the analysis in Section~\ref{section:per_package_analysis}.

\paragraph{Autovectorization by Compilers}
Autovectorization is a technique to generate vector instructions without the explicit intervention of the programmer.
For example, gcc 14 has multiple autovectorization features~\cite{gcc14_release_notes,arm_gcc14_technote} including {\it early break support},
which generates vector instructions from a loop that might exit depending on a value in memory before the proper finish condition.
Much work is also done such as quantifying the autovectorization ability of compilers~\cite{Adit2022,Carpentieri2025}
and an autovectorization framework for dynamic programming code written in Python~\cite{Moghaddasi2025}.
Although preferable for performance, autovectorization could negatively impact the safety against Downfall.
As we discuss in Section~\ref{section:per_package_analysis}, many popular packages do not require high performance
and we argue that it is better not to apply autovectorization for these packages.

\paragraph{Mitigation Measures against Downfall}
Intel provides two mitigation measures against Downfall.
Microcode is a binary blob that changes the internal logic of processors to fix bugs.
{\it Vector register scrubbing}~\cite{Intel_GDS_technote} is a software-based technique to ``scrub'' vector registers (overwrite them with random data) after use, which the programmer needs to deploy explicitly.
However, none of them completely mitigates the attack.
Microcode fixes the issue behind the curtain and the user does not need to do anything other than installing it.
The issue of microcode is the performance overhead, where {\it specific workloads may show performance impacts of up to 50\%}~\cite{Intel_GDS_technote}.
Vector register scrubbing requires recompilation of the software and a precise analysis of when to scrub.
This analysis is crucial because scrubbing after all uses of vector registers may cause huge performance overhead.

\section{Conclusion}
Downfall is a serious side-channel attack and its impact is discussed both by the original paper and by Intel to some extent.
However, its impact on daily computing activities is yet to be explored.
To this end, we extracted binary files from all packages in recent Ubuntu versions and analyzed the usage of vector registers.
Our analysis of 432~K binary files from 133~K packages found that over 60\% of all binary files use at least one vector register,
many of these usages originate from the standard libraries of C and C++,
and that some popular packages such as apt might also be affected.

\begin{acks}
This work was supported by JST, PRESTO Grant Number JPMJPR22P1, Japan.
\end{acks}

\bibliographystyle{ACM-Reference-Format}
\bibliography{main}

\end{document}